\documentclass{article}
\usepackage{spconf,amsmath,graphicx}
\usepackage{enumitem}
\usepackage[table]{xcolor}

\DeclareMathOperator*{\softmax}{SoftMax}
\title{Multi-head Cross-attentional PPG and motion signal fusion for heart rate estimation}
\makeatletter
\def\@name{
    \emph{Panagiotis Kasnesis\textsuperscript{§†}},
    \emph{Lazaros Toumanidis\textsuperscript{§†}},
    \emph{Alessio Burrello\textsuperscript{*}},\\
    \emph{Christos Chatzigeorgiou\textsuperscript{§†}},
    \emph{Charalampos Z. Patrikakis\textsuperscript{§†}},\\
}
\makeatother
\address{
	\textsuperscript{§}ThinGenious PC, Maroussi, Greece \\
	\textsuperscript{†}Department of Electrical and Electronics Engineering, University of West Attica, Greece \\
	\textsuperscript{*}Department of Electrical, Electronic and Information Engineering, University of Bologna, Italy
}
\begin{document}
\maketitle
\begin{abstract}
Nowadays, Hearth Rate (HR) monitoring is a key feature of almost all wrist-worn devices exploiting photoplethysmography (PPG) sensors. However, arm
movements affect the performance of PPG-based HR tracking. This issue is usually addressed by fusing the PPG signal with data produced by inertial measurement units. Thus, 
deep learning algorithms have been proposed, but they are considered too complex to deploy on wearable devices and lack the explainability of results. In this work, we present a new deep learning model, PULSE, which exploits temporal convolutions and multi-head cross-attention to improve sensor fusion's effectiveness and achieve a step towards explainability. We evaluate the performance of PULSE on three publicly available datasets, reducing the mean absolute error by 7.56\% on the most extensive available dataset, PPG-DaLiA. Finally, we demonstrate the explainability of PULSE and the benefits of applying attention modules to PPG and motion data.
\end{abstract}
\section{Introduction}
\label{sec:intro}
In recent years, wrist-worn devices (i.e., smartwatches) enable a 24h-monitoring of the subject's vital conditions thanks to miniaturized sensors, becoming increasingly popular in personalized health care and medical IoT applications~\cite{yeole2016use}.
One of the most important indices to monitor is Heart Rate (HR).
Compared to first-generation monitoring devices, which exploit a simple 1-3 leads ECG connected through a chest strip, modern ones use photoplethysmographic (PPG) sensors, allowing HR monitoring to be integrated into the smartwatches~\cite{troika2014}. 
However, a limitation of PPG-based HR monitoring is given by the presence of motion artifacts (MA).
These are caused by variations of sensor position on the wrist or ambient light leaking in the gap between the sensor and the wrist. 
In literature, this problem has been first tackled utilizing filtering approaches. They use the correlation between acceleration data and the PPG signal to cancel the noise and remove the MAs. Then, the HR is extrapolated from the cleaned signal~\cite{joss2015,huang2020robust}.
The critical limitation of these approaches is the high number of free hyper-parameters, which often limits their generalization.

Deep learning approaches have been proposed to improve generalization, bringing promising results on different public datasets \cite{DeepPPG2019,cornet2019, Burrello2021QPPGEP, burrello2022embedding}. 
On the other hand, these models lack explainability, since acceleration and PPG data are fused in a \emph{black box}. 
Until now, little attention has been posed to recent Transformers, given the usual high number of parameters required to train them. These models are based on the so-called \emph{Attention Modules}, which correlate different tensors.

In this paper, we demonstrate that combining feature maps of convolutions with attention modules leads to improved accuracy in the PPG-based HR monitoring and allows to interpret the connection between acceleration and PPG data.
The main contributions of this work are summarized as follows:
\begin{itemize}[leftmargin=*]
    \item We introduce a new state-of-the-art, yet lightweight (around 130M parameters),  deep neural network to fuse PPG and motion signals for precise heart rate estimation. The model includes both temporal convolutional and multi-head cross-attention modules.
    \item We evaluate the effectiveness of the produced model on three publicly available datasets. On the largest one, PPG-DaLiA \cite{DeepPPG2019}, we improve the mean absolute error (MAE) to 4.03 beats per minute (BPM), outperforming the best state-of-the-art model (a pure CNN) by 0.33 BPM.
    \item We demonstrate the explainability of the developed model and the benefits of applying attention modules to PPG and motion data by showcasing examples of attentional maps.

\end{itemize}
\begin{figure*}[btp]
\centering\includegraphics[width=\textwidth]{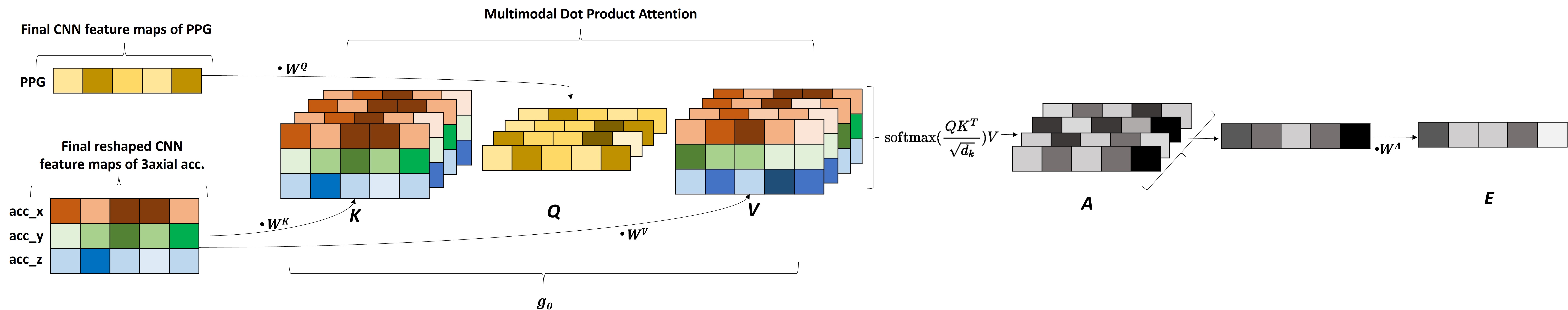}
\caption{Multi-head cross-attention module applied to PPG and 3axial accelerometer feature maps. The PPG embedding acts as a Query tensor, while the 3-axial accelerometer embeddings are Key and Value tensors.
\label{fig:cross_att}}
\vspace{-0.6cm}
\end{figure*}

\section{Background}

\label{sec:background}
\textit{Temporal Convolutional Networks:}
TCNs are 1D-Convolutional Neural Networks (CNNs), with the insertion of \textit{dilation} in convolutional layers~\cite{bai2018empirical,lea2016temporal}. The dilation is a fixed gap $d$ inserted between input samples before being convolved with the weights, thus increasing its temporal receptive field.
A convolutional layer in a TCN is formulated as:
\vspace{-0.2cm}
\begin{equation}\label{eq:1d_conv}
\mathbf{y}_t^m = \sum_{i=0}^{K-1} \sum_{l=0}^{C_{in}-1} \mathbf{x}_{t-d\,i}^l \cdot \mathbf{W}_i^{l,m}
\vspace{-0.2cm}
\end{equation}
where $\mathbf{x}$ and $\mathbf{y}$ are the input and output feature maps, $t$ and $m$ the output time-step and channel, respectively, $\mathbf{W}$ the filter weights, $C_{in}$ the number of input channels, $d$ the dilation factor, and $K$ the filter size.

\textit{Attention Module:}
An Attention Module is a neural network layer that correlates and emphasizes important parts of the input.
The classical formulation is the multi-head self-attention (MHSA)~\cite{vaswani2017attention}.
MHSA takes a tensor $\mathbf {} $ of sequential data as input and correlates it with itself.
First, it projects the sequence $\mathbf{X}$ to three separate tensors, known as \textit{queries} $\mathbf{Q}$, \textit{keys} $\mathbf{K}$ and \textit{values} $\mathbf{V}$:
\begin{equation}
    \mathbf{Q} = \mathbf{X}\mathbf{W}_\text{query} \qquad
    \mathbf{K} = \mathbf{X}\mathbf{W}_\text{key}   \qquad
    \mathbf{V} = \mathbf{X}\mathbf{W}_\text{value}
\label{eq:linear}
\end{equation}
After, $\mathbf{Q}$, $\mathbf{K}$ and $\mathbf{V}$ are used to compute the \textit{scaled dot-product attention} defined as
\begin{equation}
    \text{Attention}(\mathbf{Q}, \mathbf{K}, \mathbf{V})
    \doteq
    \mathbf{A}
    \doteq
    \softmax_\text{over keys} \left(\frac{\mathbf{Q}\mathbf{K}^\text{T}}{\sqrt{d}} \right)\mathbf{V}
\end{equation}
where $\mathbf{A}$ is the the \textit{scaled dot-product attention}, and $d$ the dimensionality of $\mathbf{K}$ used as a scaling factor.

\section{Material \& Methods}
\label{sec:material}
\subsection{Network Architecture}
\label{sec:network}
We propose the use of Multi-Head Cross-Attentional (MHCA) module for sensor modality fusion (Fig. \ref{fig:cross_att}). MHCA in contrast to MHSA, which correlates the input tensor $\mathbf{X}$ with itself, has as input two tensors $\mathbf{X_1}$ and $\mathbf{X_2}$; one of them represents the \textit{key} $\mathbf{K}$ and \textit{value} $\mathbf{V}$ tensors, while the other the \textit{query} $\mathbf{Q}$ tensor \cite{Li2022DeepFusionLD}.
In our work, these tensors correspond to different modalities (i.e., PPG and 3axial accelerometer-based feature maps). 

\begin{figure}[htb]
\begin{minipage}[b]{1.0\linewidth}
  \centering
  \centerline{\includegraphics[width=6.5cm]{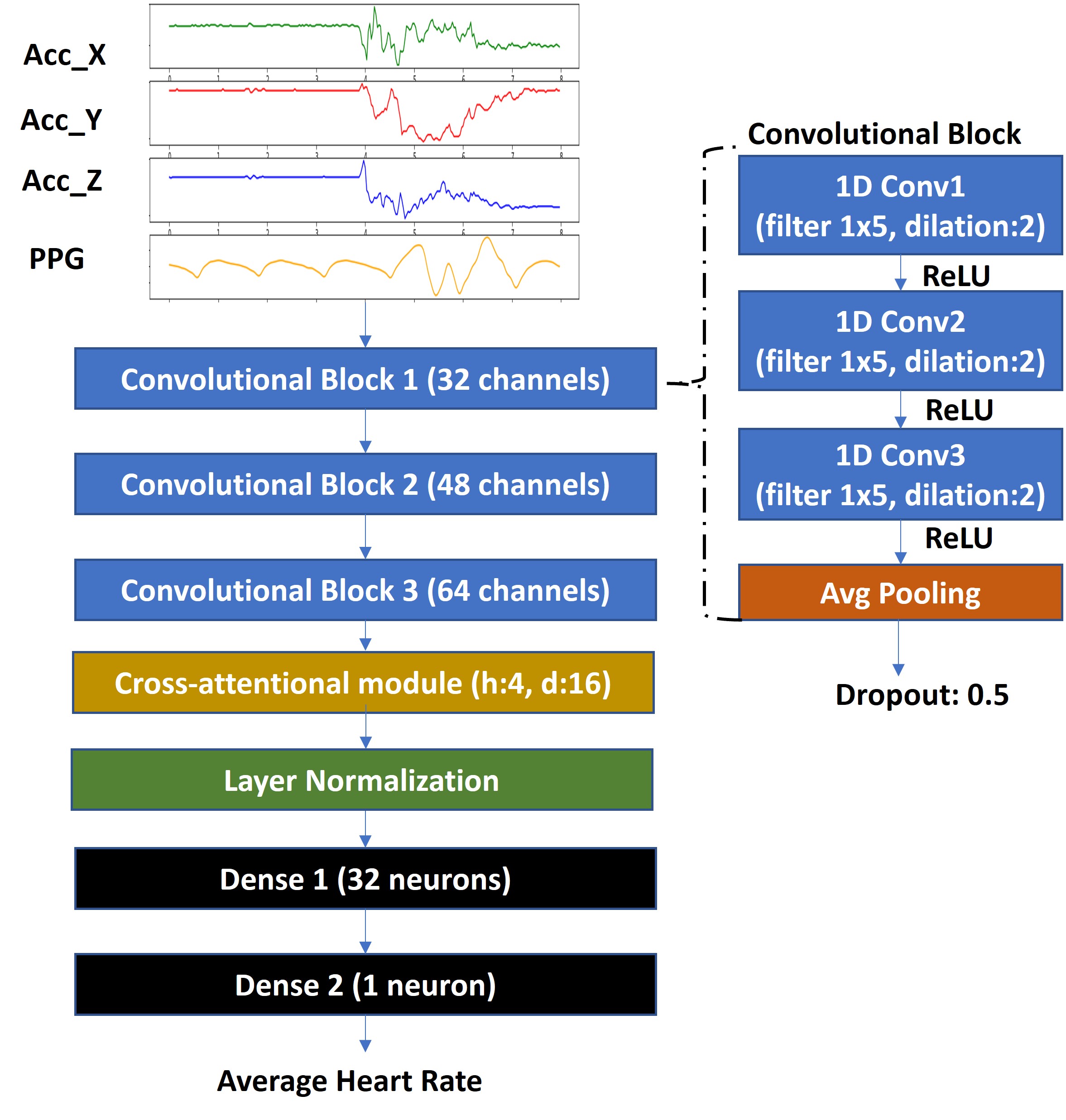}}
\end{minipage}
\caption{PULSE network architecture}
\label{fig:arch}
\vspace{-0.4cm}
\end{figure}

The proposed model architecture (selected after hyperparameter optimization) is called PULSE (\textbf{P}pg and im\textbf{U} signa\textbf{L} fu\textbf{S}ion for heart rate \textbf{E}stimation) and is displayed in Fig.\ref{fig:arch}. Three 1D convolutional blocks process the input tensor with increasing channels (i.e., 32, 48 and 64). Each block comprises 3 consecutive dilated convolutions (similar to \cite{Burrello2021QPPGEP}). The selected dilation rate was equal to 2. Afterward, the produced feature maps are fed to the MHCA module (Fig \ref{fig:cross_att}). The PPG acts as a Query vector $\mathbf{Q}$, while the 3axial accelerometer embeddings as Key $\mathbf{K}$ and Value $\mathbf{V}$ vectors. Thus, for $h$ heads, the cross-attention is computed, and, afterward, the $h$ dot products (i.e, 4 for our case) are concatenated and transformed into $\mathbf{E}$ using a dense layer again. 

\begin{table*}[hbtp] 
\caption{Per subject MAE performance of PULSE on the PPG-DaLiA dataset compared to state-of-the-art algorithms.\label{tab2}}

\begin{center}
\footnotesize
\begin{tabular}{l|lllllllllllllll|l}
\hline
Model	& S1 & S2 & S3 & S4 & S5 & S6 & S7 & S8 & S9 & S10 & S11 & S12 & S13 & S14 & S15 & Mean\\
\hline
\multicolumn{4}{c}{Deep learning models}\\
\hline
\cite{Reiss2019DeepPL} & 7.73 & 6.74 & 4.03 & 5.90 & 18.51 & 12.88 & 3.91 & 10.87 & 8.79 & 4.03 & 9.22 & 9.35 & 4.29 & 4.37 & 4.17 & 7.65\\

\cite{Song2021NASPPGPH}&5.46&5.01&3.74&6.48&12.68&10.52&3.31&8.07&7.91&	3.29&7.05&6.76&3.84&4.85&3.57&6.02\\

\cite{Burrello2021QPPGEP}&4.29&3.62&2.44&5.73&10.33&5.26&\textbf{2.00}&7.09&8.60&3.09&4.99&6.25&1.92&3.02&3.55&4.81\\

\cite{Burrello2021QPPGEP}*&\textbf{3.78}&3.36&2.33&4.84&9.95&4.38&2.20&\textbf{5.88}&7.59&2.74&4.55&5.20&2.14&\textbf{2.99}&3.47&4.36\\

Ours&4.75&3.31&2.22&5.25&7.43&4.22&2.28&8.93&6.95&2.93&3.98&6.57&\textbf{1.70}&3.22&2.88&4.44\\

Ours*&\textbf{3.78}&\textbf{3.04}&\textbf{2.20}&\textbf{4.41}&6.95&3.71&2.39&8.17&\textbf{6.19}&\textbf{2.60}&3.85&5.22&1.98&3.13&\textbf{2.79}&\textbf{4.03}\\
\hline

\multicolumn{4}{c}{Mathematical models}\\
\hline
\cite{Schck2017ComputationallyEH}&33.1&27.8&18.5&28.8&12.6&8.7&20.65&21.8&22.3&12.6&21.1&22.8&27.7&12.1&16.4&20.5\\

\cite{Salehizadeh2016ANT}&8.86&9.67&6.40&14.1&24.06&11.34&6.31&11.25&16.04&6.17&15.15&12.03&8.50&7.76&8.29&11.06\\

\cite{Huang2020RobustPA}&4.50&4.50&3.20&6.00&5.00&3.40&2.80&6.30&8.00&2.90&5.10&\textbf{4.70}&3.10&5.00&4.10&4.57\\

\cite{Zhou2020HeartRM}&5.40&4.30&3.00&8.00&\textbf{2.20}&\textbf{2.80}&3.30&8.50&12.60&3.60&\textbf{3.60}&6.10&3.00&5.50&3.70&5.04\\

\hline
\end{tabular}
\end{center}
\vspace{-0.7cm}
\end{table*}

The proposed MHCA-based fusion module operates on human actions, i.e., high-level features extracted by the temporal convolutions applied to short-term variations of the 3axial acceleration and the arterial translucency \cite{Kasnesis2021ModalitywiseRR}. Such an action maybe an arm forth movement (walking activity) or steering the wheel (driving activity). Thus, MHCA takes as input a set of action embeddings $e$ = $\{e_1^1, e_2^1...,e_1^2,e_2^2...,e_n^m\}$ extracted from signal values included in a time window (i.e., 32 actions for each modality are produced in an 8-second window), where $e_n^m$ is the $n$-th action embedding measured by the $m$-th sensor modality; MHCA is responsible for discovering the relationships between the action embeddings measured by the 3axial accelerometer conditioned by those measured by the PPG sensor, while for the case of MHSA they would be also conditioned by themselves (increasing the complexity to quadratic \cite{Chen2021CrossViTCM}). After applying MHCA the $\mathbf{E}$ tensor is produced, which is normalized and passed through 2 dense layers, with the last one producing the estimated heart rate.

\subsection{Datasets}
\label{sec:datasets}
\textit{PPG-DaLiA:}
\label{sssec:PPG-Dalia}
The first dataset used for evaluation is PPG-DaLiA, a public dataset including physiological and motion data for PPG-based heart rate estimation. The dataset was collected using a wrist-worn placed on the subjects’ non-dominant wrist to produce the one-channel PPG (sampling rate 64Hz) and the 3axial accelerometer (sampling rate 32Hz) signals while they wear a smart belt on their chest for the ground truth ECG labels. Fifteen subjects aged 21–55 years, are included in the dataset. It is worth noticing that the subjects performed a wide range of activities under close to real-life conditions, which are included as extra labels for human activity recognition \cite{Hwang2022HierarchicalDL}. The dataset includes 64,697 samples after segmentation (8-second window with 6-second overlap).\\
\textit{IEEE\_Training \& IEEE\_Test:}
\label{sssec:IEEE_Training}
These datasets are recorded using a wrist-worn device to produce a two-channel PPG sensor and a three-axis accelerometer. The sampling rate of all signals is 125 Hz. In \textit{IEEE\_Training}, twelve subjects participated (aged 18–35 years), performing a running activity at varying speeds. In \textit{IEEE\_Test},  8 different subjects (aged 19–58 years) participated while performing arm rehabilitation exercises or intensive arm movements (such as boxing). In addition, the datasets provide the average heart rate on 8-second sliding windows (window shift: 2 seconds) extracted from the raw ECG signal to be used as ground truth. The datasets include 1,768 samples (\textit{IEEE\_Training}) and 1,328 samples \textit{IEEE\_Test}.
\section{Results}
\label{sec:typestyle}
We used Python 3.8 and the PyTorch framework to design and train the neural network. 
We validated all models following the cross-validation protocol proposed in \cite{Reiss2019DeepPL, Burrello2021QPPGEP}, denoted as Leave-One-Session-Out (LOSO) cross-validation, where the 15 subjects are divided in four data folds; 3 are used as training set, while the remaining one is subdivided to form the test set (1 subject), and the validation set. This approach leads to 15 training iterations ensuring the generalizability of the produced model to unseen subject data.
Both the datasets are downsapled to 32 Hz, using input windows of $4\times256$ (PPG-DaLiA) and $5\times250$ (IEEE datasets). The windows are then normalized using per-channel z-score. 
We selected Adam \cite{Kingma2015AdamAM} as network optimizer, having the following hyperparameters: learning rate equal to 0.0005 ($\beta_1$:0.9, $\beta_2$:0.999, $\epsilon_1$:1e-08). Moreover, the batch size is set to 256 and the number of epochs to 500, with a patience of 150. The validation model that had the lowest error is used test set.
Finally, we followed the same post-processing method described in \cite{Burrello2021QPPGEP}, where the output values are clipped in case the prediction is more or less 10\% of the averaged 10 last estimated values.

\subsection{Results on PPG-DaLiA}
\label{sssec:dalia}

Table \ref{tab2} presents the MAE results produced by PULSE compared with state-of-the-art (SOTA) deep learning-based and mathematical models, with $^*$ symbol denoting the use of post-processing. Our approach achieved an average MAE equal to 4.44 BPM, reduced to 4.03 BPM when using post-processing. For most subjects, PULSE surpassed
the SOTA performance of both classical and deep learning models. Furthermore, the worse MAE obtained across patients is 8.17 BPM, which still allows for a good assessment of the health conditions. 

\begin{figure}[htb]
  \centering{\includegraphics[width=5.5cm]{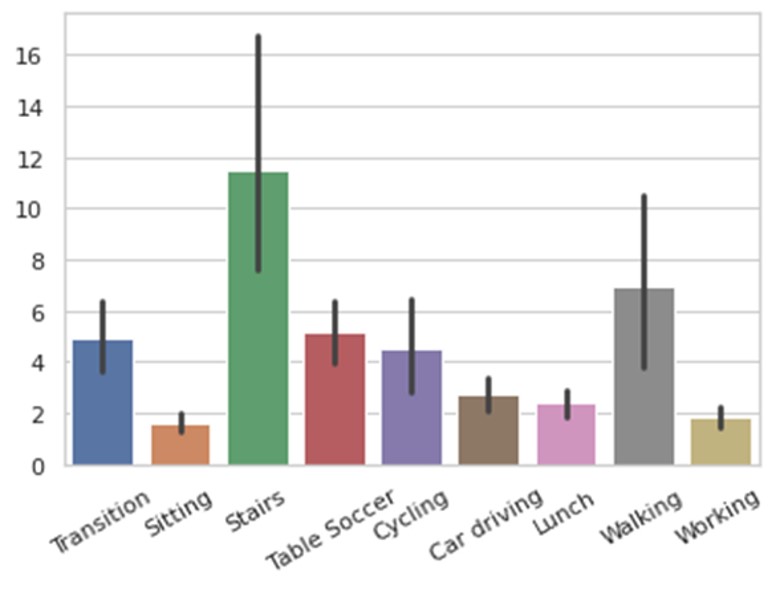}}
%
\caption{MAE results per activity on PPG-DaLiA.}
\label{fig:act}
\vspace{-0.7cm}
\end{figure}

We also performed an error analysis (MAE) w.r.t. the activity classes. As expected and shown in Fig. \ref{fig:act}, the static activities (e.g., working in an office) are more robust than the motion activities (e.g., walking) which produce higher MAE. The lowest error is achieved when the subjects performed the sitting activity and the highest when they are ascending/descending stairs.

Table \ref{modalities} discusses different options for input data and attention modality, switching between PPG only and PPG + acceleration as input, and varying between all the possible combinations in the attention module. As can be observed, the PULSE architecture is not only the most effective one, but also the most lightweight in terms of parameters. Using only the PPG as input and the MHSA module, lead to the same amount of parameters but a higher MAE of 4.52 BPM. Modifying the MHCA module and assigning the PPG to the K and V vectors, instead, increase both the MAE to 4.44 BPM and the number of parameters to 197.36k. We want also to underline that the  self-attention-based network having as input PPG and accelerometer signals is the larger architecture and it achieves the best MAE on many subjects, but does not generalize across some others (e.g., subject 5), causing a worse overall average MAE.
Our intuition is that using the MHCA module allows for a lower overfitting of the dataset, guiding the network to exploit correlations similar to the ones of SOTA classical algorithms.

\begin{table}[ht] 
\caption{Ablation study on input sensors \& attention types.
\label{modalities}}
\begin{center}
\begin{tabular}{l|l|l|l}
\hline
\textbf{Modality}	& \textbf{Attention} & \textbf{MAE}& \textbf{\#KParams}\\
\hline
PPG & MHSA & 4.52 & \textbf{131.82}\\
PPG, Acc& MHSA & 5.11 & 230,13 \\
PPG (Q) , Acc (K,V) & MHCA & \textbf{4.03}& \textbf{131.82}\\
PPG (K,V) , Acc (Q) & MHCA & 4.44 & 197.36\\

\hline
\end{tabular}
\end{center}
\vspace{-0.9cm}
\end{table}

\subsection{Results on IEEE datasets}
\label{sssec:ieee}

Table \ref{ieeetab} presents the obtained results of the developed PULSE model in comparison with the DeepPPG \cite{Reiss2019DeepPL} for the cases of the $IEEE\_Train$ and $IEEE\_Test$ datasets. In the former dataset, DeepPPG achieved lower MAE, but we should mention that this model is a huge ensemble of networks, consisting of 7 deep convolutional models, each one of them having 8.5M parameters, whilst our has around 130K parameters. So, a good trade-off between performance and complexity is demonstrated using PULSE. On the other hand, PULSE obtained the best performance when applied to $IEEE\_Test$ dataset. It is worth noticing that apart from having two PPG channels as input instead of one, no modification and hyper-parameter tuning are applied to PULSE (tuned on PPG-DaLiA), which indicates its generalizability.

\begin{table}[htbp] 
\caption{MAE performance of PULSE on the IEEE datasets
\label{ieeetab}}
\begin{center}
\begin{tabular}{l|l|l}
\hline
\textbf{Model}	& \textbf{IEEE\_Train} & \textbf{IEEE\_Test} \\
\hline
DeepPPG \cite{Reiss2019DeepPL} & \textbf{4.00} & 16.51\\
\hline
Ours&5.03&16.54\\
Ours*&4.42&\textbf{13.53}\\
\hline
\end{tabular}
\end{center}
\vspace{-0.7cm}
\end{table}

\subsection{Model explainability}
\label{sssec:XAI}

\begin{figure}[htb]
\centering{\includegraphics[width=8.7cm]{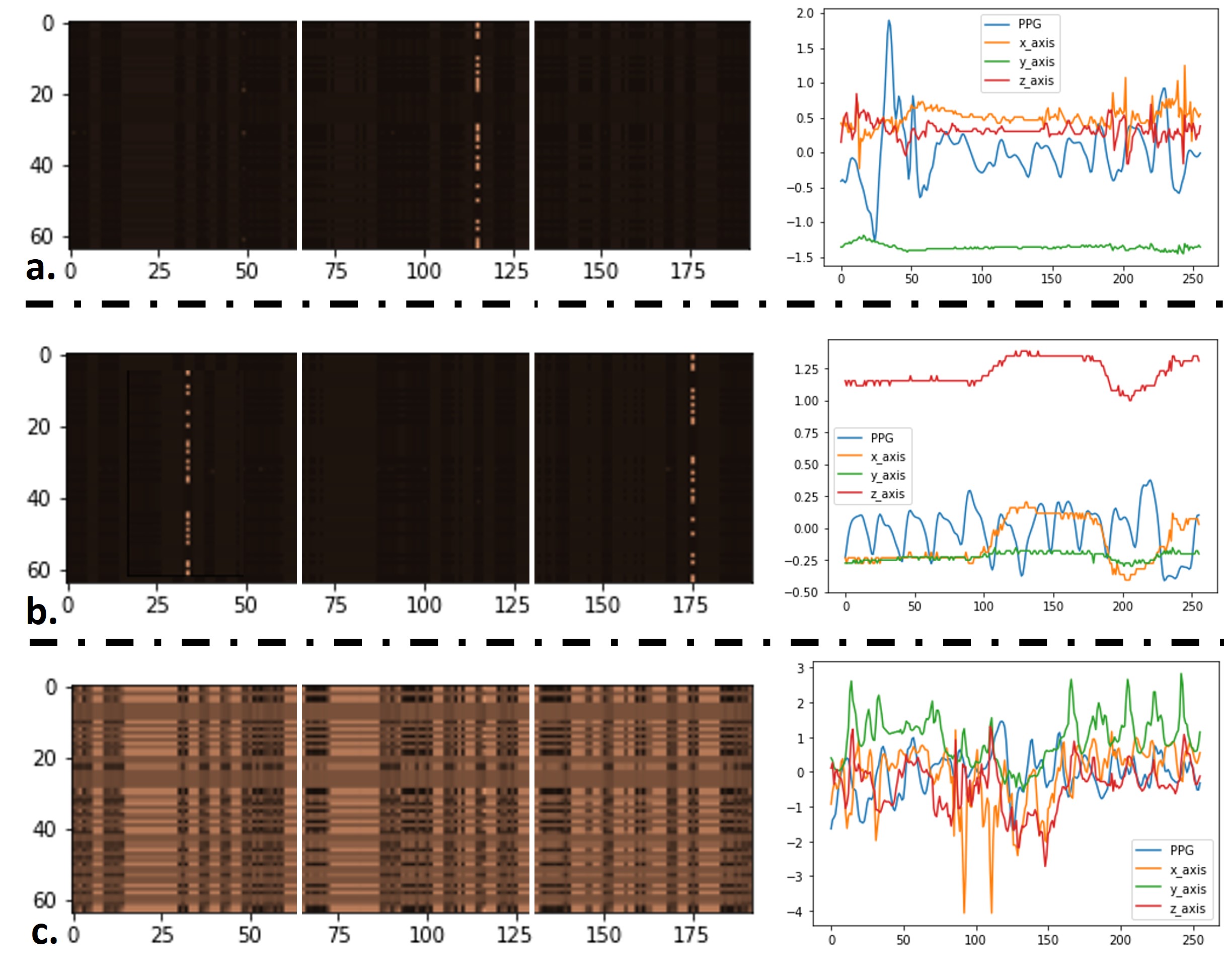}}
%
\caption{Example of the produced attentional maps. The activities performed are: a. lunch, b. walking, and c. transition.}
\label{fig:att_maps}
\vspace{-0.7cm}
\end{figure}

Apart from the low obtained MAE, one of the benefits of using an attention-based layer is that of visualizing the attentional maps to achieve explainability. Fig. \ref{fig:att_maps} presents few examples and the corresponding pre-processed input signals. While there has been no defined explainability metric for wearable-based heart rate estimation, it is useful to check in which accelerometer axis the PPG signals pay more attention (shown in dark orange). The attentional map is divided, using white horizontal lines, into three 64x64 attentional maps, i.e., one for each accelerometer axis (x, y, and z) combined with the PPG. Thus, in Fig. \ref{fig:att_maps}a. we observe that for the displayed “lunch” example, the model pays more attention to the y-axis (even though it has the least oscillations), since y-axis is usually the more stressed while eating. Fig. \ref{fig:att_maps}b. displays that for the selected “walking” example, the most important values belong to the x and z-axis, which share a similar pattern, given that their values are associated to the oscillations of the superior arms. Finally, Fig. \ref{fig:att_maps}c (transition) shows that, as expected, the model’s output is not based on a particular axis 
(input signals have many oscillations).

\section{CONCLUSION}
\label{sec:prior}

In this work, we propose a deep neural network based on a cross-attention module to fuse physiological and motion signals for heart rate detection. The method achieved 4.03 BPM as MAE on the PPG Dalia dataset, outperforming state-of-the-art algorithms, and generalize over multiple datasets, achieving state-of-the-art results on a benchmarked dataset. Further, we show how the attention maps can be exploited for explainability of the model. 

\vfill\pagebreak

\bibliographystyle{IEEEbib}
\bibliography{refs}

\end{document}